\documentclass[aps,prl,reprint,superscriptaddress,floatfix]{revtex4-2}
\usepackage{graphicx}
\usepackage{dcolumn}
\usepackage{bm}
\usepackage{amsmath}
\usepackage{overpic}
\usepackage{subfigure}
\usepackage{ragged2e}
\usepackage{amssymb}

\usepackage{braket}

\usepackage[colorlinks=true,citecolor=blue,urlcolor=blue]{hyperref}

\begin{document}



\title{Tunable Exceptional Points for Quantum Sensing in a Spin--Orbit-Angular-Momentum Coupled BEC}


\author{Zicheng Zhang}
\affiliation{Laboratory of Quantum Information, University of Science and Technology of China, Hefei 230026, China}
\affiliation{Anhui Province Key Laboratory of Quantum Network, University of Science and Technology of China, Hefei 230026, China}
\author{Xi-Wang Luo}\email{luoxw@ustc.edu.cn}
\affiliation{Laboratory of Quantum Information, University of Science and Technology of China, Hefei 230026, China}
\affiliation{Anhui Province Key Laboratory of Quantum Network, University of Science and Technology of China, Hefei 230026, China}
\affiliation{CAS Center For Excellence in Quantum Information and Quantum Physics, University of Science and Technology of China, Hefei 230026, China}
\affiliation{Hefei National Laboratory, University of Science and Technology of China, Hefei 230088, China}
\affiliation{Anhui Center for Fundamental Sciences in Theoretical Physics, University of Science and Technology of China, Hefei, 230026, China}
\date{\today}

\begin{abstract}
Exceptional points (EPs) can induce strongly amplified responses to weak perturbations, but enhanced spectral sensitivity in conventional non-Hermitian systems does not necessarily translate into improved quantum-limited sensing because of the associated gain and loss noise. Here, we investigate a tunable EP sensing platform based on the intrinsic non-Hermitian Bogoliubov dynamics of a spin-orbit-angular-momentum-coupled Bose-Einstein condensate.
Starting from a fully Hermitian microscopic Hamiltonian, we derive a Bogoliubov dynamical matrix that exhibits parity-time (PT) symmetry, with EPs tunable through the Raman coupling and interaction parameters. We identify multiple EPs and map their trajectories and associated stability landscapes, revealing strong quantum Fisher information enhancement when the EPs are approached from the PT-symmetric stable regime. 
We further find that the gap-opening rate around a second-order EP provides a useful relative indicator for comparing the sensing performance of EPs, while higher-order EPs are not necessarily accessible from the stable regime. Moreover, two second-order EPs can be tuned to overlap, allowing their sensing contributions to add and yielding a linear enhancement of the total quantum Fisher information. Finally, we show that a spin-density measurement can approach the quantum Fisher information limit, providing an experimentally accessible route to tunable EP-enhanced quantum sensing in Bose-Einstein condensates.

\end{abstract}

\maketitle

\section{Introduction}
Quantum metrology seeks to exploit quantum resources to improve the precision of parameter estimation and has become an important component of quantum information science and technology~\cite{Braunstein1994StatisticalDistance,Giovannetti2004QuantumEnhanced,Paris2009QuantumEstimation,Giovannetti2011AdvancesMetrology,Degen2017QuantumSensing}. A promising route to enhanced sensing is to exploit criticality, where the response to a weak perturbation can become singular or strongly amplified near a critical point~\cite{Chu2021CriticalityEnhanced,Ding2022EnhancedMetrology}. Exceptional points (EPs), at which both eigenvalues and eigenvectors of a non-Hermitian Hamiltonian coalesce, provide a paradigmatic example of such critical behavior~\cite{Bender1998RealSpectra,Heiss2012PhysicsEP,Bergholtz2021ExceptionalTopology,Ashida2020NonHermitianPhysics,ElGanainy2018NonHermitianPT,Ozdemir2019PTExceptionalPhotonics,Miri2019ExceptionalOptics}. The nonanalytic spectral response near an EP has motivated extensive studies of EP-based sensing and demonstrated strongly enhanced spectral responses to weak perturbations~\cite{Ruter2010ObservationPT,Wiersig2014EnhancedSensitivityEP,Chen2017ExceptionalSensingMicrocavity,Hodaei2017HigherOrderSensitivity}. However, enhanced spectral splitting does not by itself imply enhanced quantum-limited sensitivity. In conventional dissipative non-Hermitian systems, the same gain or loss processes responsible for the EP response also introduce quantum and thermal fluctuations, which can compensate for, or even overwhelm, the apparent signal enhancement~\cite{Langbein2018NoExceptionalPrecision,Lau2018FundamentalLimits,Zhang2019QuantumNoiseEP,Chen2019SensitivityParameterEstimation,Wang2020PetermannLimit,Ding2023FundamentalSensitivity,Loughlin2024NoSNRAdvantage,Almanakly2026WaveguideQED}. Thus, a central challenge is to realize EP-induced amplification while avoiding the noise penalty inherent to conventional gain-loss implementations.

To overcome this limitation, recent studies have shown that nonlinear exceptional singularities can exhibit sensing characteristics that differ qualitatively from their linear counterparts, potentially leading to enhanced signal-to-noise performance~\cite{Bai2023NonlinearityHigherOrder,Zheng2025NoiseNonlinearEP}. Another particularly attractive route is to realize effective non-Hermitian dynamics within an underlying Hermitian quantum system~\cite{McDonald2020ExponentialQuantumSensing,Chu2020PseudoHermitianQubitSensing,Naikoo2023MultiparameterNonHermitian,Yu2024HeisenbergNonHermitian,Xiao2024SensingWithoutEP,Arkhipov2026QFIBoundPseudoHermitian,Wang2026SqueezingEnhancedEP,Ding2021TrappedIonEP}. In this setting, the Hamiltonian remains Hermitian and no external gain or loss reservoir is required, while the corresponding quantum dynamics can nevertheless exhibit exceptional points. Such an intrinsic realization can therefore access non-Hermitian critical dynamics without the Langevin-type noise associated with dissipative gain and loss. 
In particular, quantum optical squeezing and dynamically unstable Bose-Einstein condensates (BECs) provide natural platforms for realizing parity-time (PT)-symmetric Bogoliubov dynamics with EP-induced amplification, which can be exploited for quantum sensing through the associated squeezing response, and higher-order EPs may offer enhanced sensitivity scaling~\cite{Peng2016AntiPTFlyingAtoms,Zhang2020SyntheticAntiPT,Nair2021AntiPTSensing,Wang2022QFIAntiPT,Luo2022PseudoAPT,Liu2024QFIScalingQEP}.

A particularly simple realization of this approach is provided by the dynamical instability of a single-component BEC, where the relevant Bogoliubov dynamics can be reduced to an effective two-mode squeezing model and exhibits an exceptional point~\cite{Luo2022PseudoAPT}. Although this provides a natural platform for EP-based quantum sensing, the relevant effective mode pair is essentially fixed by the single-particle dispersion, leaving very limited freedom to tune the EP location and the associated sensing conditions. A natural way to overcome this restriction is to engineer the single-particle dispersion itself. Spin-orbit-angular-momentum (SOAM)-coupled BECs provide an attractive platform for this purpose~\cite{Lin2011SpinOrbitBEC,Sun2015SOAMBEC,Qu2015QuantumPhasesSOAM,Chen2018SOAMExperiment,Zhang2019SOAMPhaseDiagram,Chen2020SOAMExcitationSpectrum,Li2013SuperstripesSOC}. 
In a ring geometry, Raman coupling between two hyperfine states transfers orbital angular momentum to the atoms and couples their internal spin to discrete angular-momentum modes~\cite{Marzlin1997VortexCoupler,Andersen2006QuantizedRotation,Ryu2007PersistentFlowToroidal,Kanamoto2007RamanLaguerreGaussian}.
The Raman coupling and two-photon detuning therefore provide direct control over the single-particle angular-momentum dispersion, allowing different angular-momentum fluctuation modes to participate in the Bogoliubov dynamics. Despite these advantages, a systematic understanding of how such tunable multimode dynamics can be exploited for exceptional-point quantum sensing is still lacking.

In this work, we address this problem by systematically investigating a tunable EP sensing platform based on a ring-trapped SOAM-coupled BEC. Starting from the underlying Hermitian many-body Hamiltonian, we derive the time-independent Bogoliubov--de Gennes (BdG) dynamical matrix describing the coupled spin and angular-momentum fluctuation modes. We show that the resulting dynamics exhibit PT symmetry and support multiple exceptional points of different orders, including second- and fourth-order EPs, whose locations can be tuned by the Raman and interaction parameters.
We map the trajectories of these EPs and the associated stability landscape, and evaluate the quantum Fisher information (QFI) for parameter estimation as the EPs are approached from the PT-symmetric stable side. We find strong sensitivity enhancement due to EP-induced dynamical amplification and 
identify the gap-opening rate around a second-order EP as a useful indicator for comparing the sensing performance. Although fourth-order EPs are present, they cannot be directly exploited for sensing 
because they are located within PT-broken unstable regimes that are separated from the stable regimes by intervening second-order EPs.
This highlights that EP accessibility and the surrounding stability landscape are as important as EP order for practical quantum sensing~\cite{Liu2024QFIScalingQEP}. We further show that two second-order EPs can be tuned to overlap, enabling independent EP channels to contribute additively to the total QFI and resulting in a linear enhancement of the sensing capability. Finally, we propose a spin-density-based measurement protocol and demonstrate that its sensitivity can approach the QFI limit, providing an experimentally accessible route to tunable EP-enhanced quantum sensing.

\section{Model and Symmetry Analysis}
\begin{figure}[tb]
    \centering
    \hspace*{-0.07\linewidth}
    \includegraphics[width=1.17\linewidth]{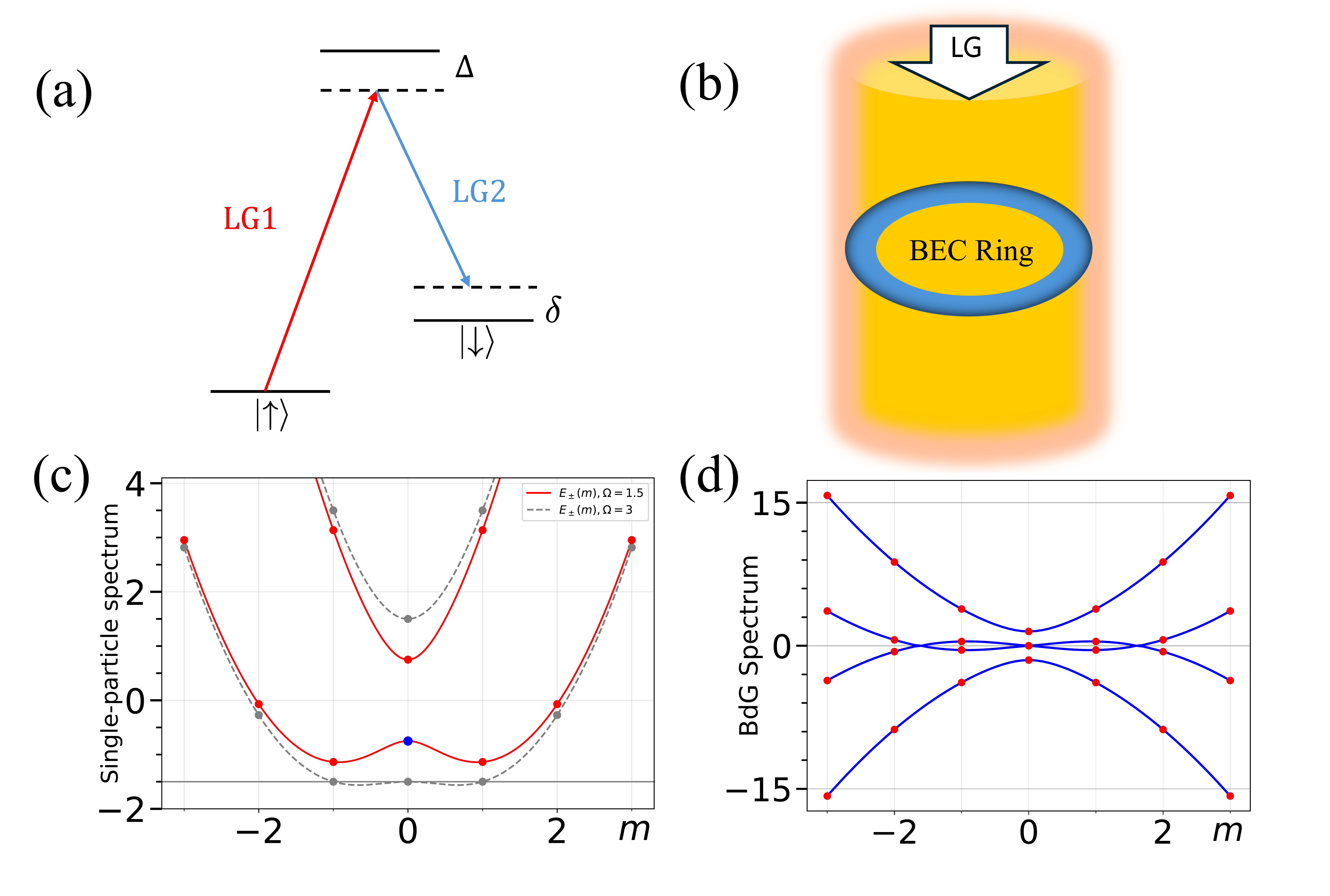}
    \caption{(a) Raman coupling between the two internal states of the BEC. (b) Schematic illustration of the ring geometry and Raman beams. (c) Single-particle energy spectrum for $\Omega=1.5$ (red) and $\Omega=3$ (gray dashed), with $l=1$ and the relevant modes indicated by dots. (d) BdG spectrum of a BEC condensed in the mode marked by the blue point in (c); the spectrum is dynamically stable with all eigenvalues real.}
    \label{fig:fig1}
\end{figure}

We consider a two-component Bose--Einstein condensate (BEC) confined in a quasi-one-dimensional ring trap of radius $R$ and azimuthal angle $\phi$~\cite{Bloch2008ManyBodyUltracold}. The two internal states, denoted by $\lvert \uparrow \rangle$ and $\lvert \downarrow \rangle$, are coupled by a pair of Raman laser beams carrying opposite orbital angular momenta. A Raman-induced spin flip is therefore accompanied by the transfer of orbital angular momentum between the two spin components. The corresponding level configuration and a schematic representation of the ring geometry are shown in Figs.~\ref{fig:fig1}(a) and \ref{fig:fig1}(b), respectively.
After applying the unitary transformation $\hat{\Psi}_{\uparrow/\downarrow} \rightarrow e^{\mp il \phi} \hat{\Psi}_{\uparrow/\downarrow}$ to the field operators, adopting the natural energy unit $\hbar^2/(2MR^2)$ and setting $\hbar=1$, we obtain 
the effective single-particle Hamiltonian in the rotating frame~\cite{Sun2015SOAMBEC}: 
\begin{equation}
    H_0 = -\partial_\phi^2 + \left(2il\partial_\phi + \frac{\delta}{2}\right)\sigma_z + \frac{\Omega}{2}\sigma_x,
\end{equation}
where $\Omega$ and $\delta$ denote the normalized Raman coupling strength and two-photon detuning, respectively, while $\sigma_{x,z}$ are the Pauli matrices acting in the spin space. 
We now consider $s$-wave scattering interactions, including the intra-spin interactions $g_\uparrow$ ($g_\downarrow$) between spin-up (spin-down) bosons and the interspin interaction $g_{\updownarrow}$. The resulting second-quantized Hermitian Hamiltonian can be written as
\begin{equation}
\mathcal{H} =\int_{0}^{2\pi} {\hat\Psi}^\dagger \left( H_0 + H_{\text{int}} \right) {\hat\Psi} d\phi,
\end{equation}
where
\begin{equation}
    H_{\text{int}} = \frac{1}{2}
\begin{pmatrix}
g_{\uparrow} {\hat\Psi}_{\uparrow}^\dagger{\hat\Psi}_{\uparrow} & g_{\updownarrow} {\hat\Psi}_{\downarrow}^\dagger {\hat\Psi}_{\uparrow} \\
g_{\updownarrow} {\hat\Psi}_{\uparrow}^\dagger {\hat\Psi}_{\downarrow} & g_{\downarrow} {\hat\Psi}_{\downarrow}^\dagger {\hat\Psi}_{\downarrow}
\end{pmatrix}
\end{equation}
with $\hat\Psi=[\hat\Psi_\uparrow,\hat\Psi_\downarrow]^T$.

Within the Bogoliubov approximation, the field operators are decomposed into a macroscopically occupied condensate mode and fluctuations orthogonal to it~\cite{Ozeri2005BogoliubovExcitations,FetterWalecka2012ManyParticle}. We assume that the condensate is initialized to the lower-energy steady state with angular-momentum mode $m_{0}$ in the rotating frame, and
the field operators can be expanded as
\begin{equation}
    \hat{\Psi}=e^{-i\mu t}\left[e^{im_0\phi+i\frac{\pi}{4}}(\Phi+\frac{\chi}{\sqrt{2\pi}}\hat\psi_{m_0}) + \sum_{m \neq m_0} \frac{e^{im\phi}}{\sqrt{2\pi}} \hat{\psi}_{m} \right]
\end{equation}
where $\Phi=[\Phi_{\uparrow},\Phi_{\downarrow}]^T$ is the two-component condensate spinor wave function with density $\rho_0=\rho_{0\uparrow}+\rho_{0\downarrow}$ and  $\rho_{0s}= \left|\Phi_{s}\right|^{2}$. Without loss of generality, $\Omega$ can be chosen to be real and positive, in which case the relevant condensate solution $\Phi_{s}$ can be taken to be real as well. 
Here $\mu$ is the chemical potential, $m_0$ denotes the angular-momentum quantum number of the condensate mode, $\hat\psi_m=[\hat\psi_{m\uparrow},\hat\psi_{m\downarrow}]^T$ denotes the fluctuation field operators. The spinor $\chi=[-\Phi_\downarrow,\Phi_\uparrow]^T/\sqrt{\rho_0}$ represents the spin excitation in the $m_0$ sector and is orthogonal to the condensate spinor $\Phi$.

Starting from the Heisenberg equations of motion and expanding the field operators around the mean-field steady state, we obtain the linearized dynamical equations for the quantum excitation operators. The dynamics decompose into independent sectors labeled by $m$. The leading-order equations for $m\neq m_0$ are given by
\begin{equation}
    i\frac{\partial}{\partial t}
\begin{pmatrix}
{\psi}_{m\uparrow} \\
{\psi}_{m\downarrow}\\
{\psi}_{m'\uparrow}^\dagger \\
{\psi}_{m'\downarrow}^\dagger
\end{pmatrix} = H_{\rm dyn}\begin{pmatrix}
{\psi}_{m\uparrow} \\
{\psi}_{m\downarrow}\\
{\psi}_{m'\uparrow}^\dagger \\
{\psi}_{m'\downarrow}^\dagger
\end{pmatrix},
\end{equation}
with $m'=-m+2m_0$ and 
\begin{equation}
    H_{\rm dyn}=\begin{pmatrix}
 \epsilon_\uparrow & \varepsilon_\uparrow & i\kappa_\uparrow & i\zeta\\
 \varepsilon_\downarrow & \epsilon_\downarrow & i\zeta & i\kappa_\downarrow\\
  i\kappa_\uparrow^*&i\zeta^*  & -\epsilon'_\uparrow & -\varepsilon_\uparrow\\
 i\zeta^* & i\kappa_\downarrow^* & -\varepsilon_\downarrow&-\epsilon'_\downarrow
\end{pmatrix},
\end{equation}
where $\epsilon_\uparrow=(m^2-2ml+
\frac{\delta}{2}+2g_\uparrow \left |\Phi_\uparrow \right |^2+g_\updownarrow \left |\Phi_\downarrow \right |^2-\mu)$, $\varepsilon_\uparrow=\frac{\Omega}{2}+g_\updownarrow \Phi_\downarrow^*\Phi_\uparrow$, $\kappa_\uparrow = g_\uparrow\Phi^2_\uparrow$ and $\zeta=g_\updownarrow\Phi_\uparrow\Phi_\downarrow$,   $\epsilon_\downarrow=(m^2+2ml-
\frac{\delta}{2}+2g_\downarrow \left |\Phi_\downarrow \right |^2+g_\updownarrow \left |\Phi_\uparrow \right |^2-\mu)$, $\varepsilon_\downarrow=\frac{\Omega}{2}+g_\updownarrow \Phi_\uparrow^*\Phi_\downarrow$, $\kappa_\downarrow = g_\downarrow\Phi^2_\downarrow$, $\epsilon'_\uparrow=(m^2+2ml-4mm_0+4m_0^2-4m_0l+
\frac{\delta}{2}+2g_\uparrow \left |\Phi_\uparrow \right |^2+g_\updownarrow \left |\Phi_\downarrow \right |^2-\mu)$, and $\epsilon'_\downarrow=(m^2-2ml-4mm_0+4m_0^2+4m_0l-
\frac{\delta}{2}+2g_\downarrow \left |\Phi_\downarrow \right |^2+g_\updownarrow \left |\Phi_\uparrow \right |^2-\mu)$.

As we discussed previously, we can choose the gauge such that the relevant condensate solution $\Phi_{s}$ are real. In this case, the parameters in $H_{\rm dyn}$ are also real, and thus $H_{\rm dyn}$ possesses PT symmetry with 
$P=\sigma_z\otimes I_2=\text{diag}\{1,1,-1,-1\}$ and $T$ the complex conjugate operator
\begin{equation}
    [PT,H_{\rm dyn}]=0.
\end{equation}
The PT symmetry ensures that the eigenvalues of $H_{\rm dyn}$ occur in pairs related by $E\leftrightarrow E^*$.
Moreover,
we can rewrite the dynamical equation as $i\partial_t \hat V=H_{\rm BdG} \hat V$ by extending the basis to 
$ \hat V=({\psi}_{m\uparrow},{\psi}_{m\downarrow}, {\psi}_{m'\uparrow},{\psi}_{m'\downarrow},{\psi}^\dagger_{m\uparrow},{\psi}^\dagger_{m\downarrow}, {\psi}^\dagger_{m'\uparrow},{\psi}^\dagger_{m'\downarrow})^T$, where
\begin{equation}
{H}_{\text{BdG}} =
\begin{pmatrix}
\mathcal{A} & \mathcal{D} \\
-\mathcal{D}^* & -\mathcal{A}^*
\end{pmatrix},
\end{equation} 
and the off-diagonal block $\mathcal{D}$ encodes anomalous correlations, so we have 
\begin{equation}
   \mathcal{A}=\begin{pmatrix}
 \epsilon_\uparrow & \varepsilon_\uparrow & 0 & 0\\
 \varepsilon_\downarrow & \epsilon_\downarrow & 0& 0\\
  0&0  & \epsilon'_\uparrow & \varepsilon_\uparrow\\
 0 &0 & \varepsilon _\downarrow&\epsilon'_\downarrow
\end{pmatrix};\quad \mathcal{D} = \begin{pmatrix}
 0 & 0 & i\kappa_\uparrow & i\zeta\\
 0 & 0 & i\zeta & i\kappa_\downarrow\\
  i\kappa_\uparrow&i\zeta  & 0 & 0\\
 i\zeta & i\kappa_\downarrow & 0&0
\end{pmatrix}.
\end{equation}
Owing to particle-hole symmetry, the full BdG dynamical matrix satisfies
\begin{equation}
    \tau_x H_{\rm BdG}\tau_x=-H_{\rm BdG}^*,
\end{equation}
so that its eigenvalues occur in pairs related by $E\leftrightarrow -E^*$. Here $\tau_x$ is the Pauli matrix acting on the particle-hole space.
Due to this particle-hole redundancy, $H_{\rm BdG}$ contains two equivalent copies of $H_{\rm dyn}$, and it is therefore sufficient to consider $H_{\rm dyn}$ to fully characterize the spectrum and dynamics of the system. 
In particular, for $\delta=0$ and $m_0=0$, $H_{\rm dyn}$ also possesses a chiral symmetry
\begin{equation}
    \{\Gamma,H_{\rm dyn}\}=0,
\end{equation}
where $\Gamma=\sigma_y\otimes \sigma_x$ is the chiral symmetry operator. This chiral symmetry is generally broken when $\delta\neq0$ or $m_0\neq0$. 
The chiral symmetry enforces the eigenvalue pairing $E\leftrightarrow -E$ in $H_{\rm dyn}$ and originates from the symmetry between the $m$ and $-m$ modes.

For $\hat\psi_{m_0}$, we have
\begin{equation}
i\partial_t\hat\psi_{m_0}
=\epsilon_0\hat\psi_{m_0}
+\kappa_0\hat\psi_{m_0}^\dagger,
\end{equation}
where $\epsilon_0
=m_0^2-\mu+\frac{(\delta-4m_0l)\left(\rho_{0\downarrow}-\rho_{0\uparrow}\right)}{2\rho_0}+\frac{\Omega\sqrt{\rho_{0\uparrow}\rho_{0\downarrow}}}{\rho_0}
+\frac{2\rho_{0\uparrow}\rho_{0\downarrow}
\left(g_{\uparrow}+g_{\downarrow}\right)
}{
\rho_0
}+\frac{g_{\updownarrow}
\left(\rho_{0\uparrow}-\rho_{0\downarrow}\right)^2
}{
\rho_0
}$
and $\kappa_0=\frac{\rho_{0\uparrow}\rho_{0\downarrow}}{\rho_0}(g_\uparrow+g_\downarrow-2g_\updownarrow)$. 
We will focus on isotropic interactions with $g_\uparrow=g_\downarrow=g_{\updownarrow}=g$ throughout this work, although our results can be generalized to more general interaction forms. In this case, we have $\kappa_0=0$, while $\epsilon_0$ reduces to the single-particle excitation energy. Thus, the $\psi_{m_0}$ mode is always dynamically stable, and we can only consider the $m\neq m_0$ modes. 
For these $m\neq m_0$ modes, though the dynamical matrix exhibits PT symmetry, its eigenstates may spontaneously break this symmetry. For the parameters considered in this work, the PT-symmetric phase is dynamically stable with a real spectrum, whereas the PT-broken phase is dynamically unstable with a complex spectrum and the quasiparticle amplitudes exhibit exponential growth. 
These two regimes are separated by exceptional points, near which the Bogoliubov dynamics can strongly enhance the QFI, motivating their use for quantum sensing. Figs.~\ref{fig:fig1}(c) and \ref{fig:fig1}(d) show the single-particle spectrum and the corresponding BdG spectrum for $\delta=0$ and $m_0=0$, with $\Omega$ and $g$ chosen such that the system is in the stable regime.

\section{Tunable exceptional points}

We now investigate how the exceptional points evolve as the Raman coupling, two-photon detuning, and interaction strength are varied. Note that the orbital angular-momentum transfer $l$ in the SOAM coupling and the condensate angular momentum $m_0$ are fixed by the system configuration. For a given set of parameters, the condensate state and chemical potential are first determined from the Gross-Pitaevskii equations, after which the EPs are identified from the corresponding Bogoliubov dynamical matrix $H_{\rm dyn}$. 

For the condensate mode $m_0$, the mean-field equations determine the relative population and phase of the two spin components. Writing
$\Phi_{\uparrow,\downarrow}=|\Phi_{\uparrow,\downarrow}|
e^{i\vartheta_{\uparrow,\downarrow}}$, the global phase can be removed and,
for $\Omega>0$, energy minimization gives
$\vartheta_\downarrow-\vartheta_\uparrow=\pi$. It is therefore convenient to
introduce the ratio
$r=\Phi_\downarrow/\Phi_\uparrow<0$. We choose a gauge in which
$\Phi_{\uparrow,\downarrow}$ are real. For isotropic interactions, the
Gross-Pitaevskii equations give
\begin{equation}
r=
\frac{\Delta_0}{\Omega}
-\sqrt{\left(\frac{\Delta_0}{\Omega}\right)^2+1},
\end{equation}
where
\begin{equation}
\Delta_0=4m_0l-\delta.
\end{equation}
The corresponding chemical potential is
\begin{equation}
\mu=m_0^2+g\rho_0-\frac{1}{2}\sqrt{\Delta_0^2+\Omega^2}.
\end{equation}
These relations are used below to calculate the BdG spectrum and locate the EPs.
Specifically, we examine the characteristic polynomial
\begin{equation}
    P(E)=\det(EI-H_{\rm dyn}),
\end{equation}
and locate the eigenvalue degeneracies from
\begin{equation}
    P(E_{\mathrm{EP}})=0,\qquad
    \left.\partial_E P(E)\right|_{E=E_{\mathrm{EP}}}=0,
\end{equation}
followed by verification of the corresponding eigenstates.
With all other parameters fixed, the EP positions in $g$, $\Omega$, and $\delta$ can therefore be obtained from these degeneracy conditions.

We first consider the symmetric case $m_0=0$ and $\delta=0$. In this case, the excitation sectors are naturally labeled by
$n\equiv m-m_0>0$, and the single-particle spectrum is symmetric under
$m=m_0+n\rightarrow m'=m_0-n$. The Raman coupling provides direct control
over the single-particle dispersion and, consequently, over the relation between
the condensate mode and the excitation sectors. In particular, the single-particle
energies of the angular-momentum modes $m,m'=\pm n$ become degenerate with the
condensate mode at the critical Raman coupling
\begin{equation}
    \Omega_c^{(n)}=4l^2-n^2,
\end{equation}
the degenerate energy is marked by the horizontal lines in Fig.~\ref{fig:fig1}(c). This critical coupling marks a reconstruction of the EP structure in the corresponding angular-momentum sector, it
exists for $n<2l$ under the condition $\Omega>0$. It separates two qualitatively
different EP structures of a given angular-momentum sector, as discussed below.

\begin{figure}[tb]
    \centering
    \hspace*{-0.07\linewidth}
    \includegraphics[width=1.0\linewidth]{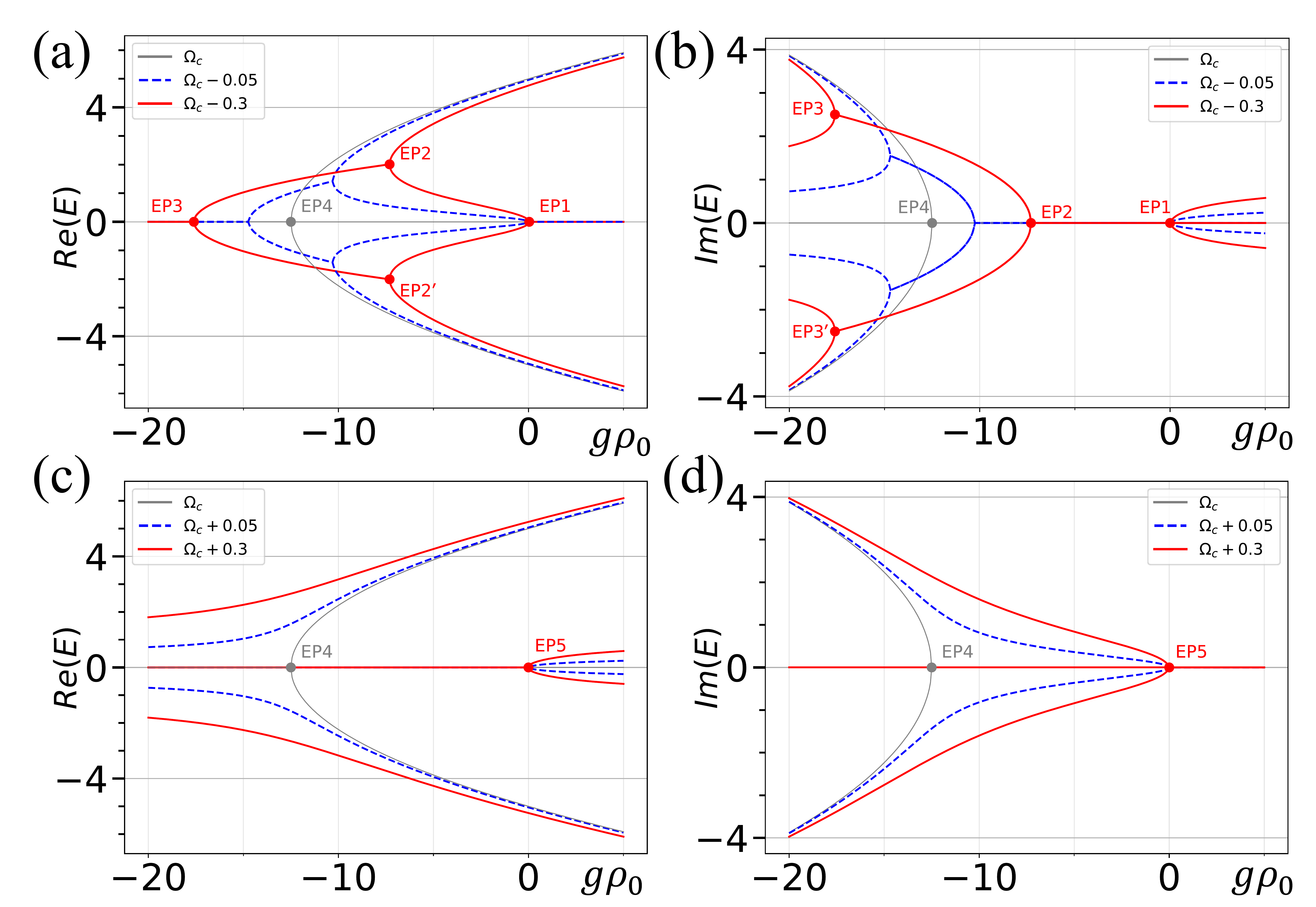}
    \caption{Evolution of the BdG spectrum across the critical Raman coupling for $n=1$, $l=1$, $m_{0}=\delta=0$, with $\Omega_{c}=3$.
(a) Real and (b) imaginary parts of the BdG eigenvalues for
$\Omega=2.95$ (blue dashed),
$\Omega=2.7$ (red solid), and
$\Omega=3$ (gray).
(c) Real and (d) imaginary parts of the BdG eigenvalues for
$\Omega=3.05$ (blue dashed),
$\Omega=3.3$ (red solid), and
$\Omega=3$ (gray).
The corresponding EPs are indicated by the labels.
}
    \label{fig:fig2}
\end{figure}

For illustration, Fig.~\ref{fig:fig2} shows the evolution of the BdG spectrum as
a function of $g$ for $l=1$, $m_0=\delta=0$, and sector $n=1$, for which
$\Omega_c^{(1)}=3$. For $\Omega<\Omega_c^{(1)}$ [see Fig.~\ref{fig:fig2}(a) and \ref{fig:fig2}(b)], the spectrum exhibits one
isolated second-order EP at $g>0$, denoted by EP1, together with two pairs of
coincident second-order EPs in the attractive-interaction regime. We denote these
two pairs by $(\mathrm{EP2},\mathrm{EP2}')$ and
$(\mathrm{EP3},\mathrm{EP3}')$, respectively, and refer to each pair as a
double-EP structure. The two double-EP structures have distinct spectral
signatures. At EP2, two pairs of eigenvalues coalesce at opposite finite values
of $\mathrm{Re}(E)$ while their imaginary parts vanish. The two
symmetry-related EPs are therefore visible as two branch-merging points in the
real spectrum. At EP3, in contrast, the real parts of the four relevant
eigenvalues coalesce at zero, while the imaginary spectrum exhibits two
symmetry-related branch-merging points at opposite nonzero values of
$\mathrm{Im}(E)$. 
The coincidence of the second-order EPs follows from the chiral symmetry,
which originates from the $m\leftrightarrow -m$ symmetry of the single-particle
dispersion. Thus, although each degeneracy is second order, two symmetry-related
EPs occur at the same parameter point. The system is dynamically stable in the
region between EP1 and EP2, where the eigenstates of $H_{\rm dyn}$ preserve
the PT symmetry. Outside this region, the PT symmetry is spontaneously broken
and the spectrum becomes complex.

As the Raman coupling is increased toward $\Omega_c^{(1)}$, the two double-EP
structures move toward each other, while EP1 moves from positive interaction
strength toward $g=0$. At the critical coupling
$\Omega=\Omega_c^{(1)}$, the two double-EP structures merge at
$g\rho_0=-12.5$, where four eigenvalues and their corresponding eigenvectors
coalesce to form a fourth-order EP, denoted by EP4. At the same critical
coupling, the dynamical matrix $H_{\rm dyn}$ develops an exceptional line at
zero energy as a function of the interaction strength. Specifically, for
$g\neq0$, the zero-energy degeneracy is a second-order EP, whereas at
$g\rho_0=-12.5$ it becomes a fourth-order EP. The point $g=0$ is an ordinary
two-fold degeneracy rather than an exceptional point. 
Thus, the fourth-order EP appears as an order-enhanced point embedded in a continuous exceptional structure rather than as an isolated singularity.

For $\Omega>\Omega_c^{(1)}$ [see Fig.~\ref{fig:fig2}(c) and \ref{fig:fig2}(d)], the fourth-order EP disappears and the
exceptional-point structure is continuously reconstructed into a single
second-order EP branch, denoted by EP5. 
As $\Omega$ changes across
$\Omega_c^{(1)}$, EP1 and EP5 approach $g=0$ from opposite sides and merge into
the ordinary degeneracy at $g=0$ when $\Omega=\Omega_c^{(1)}$. 
Although EP5 is continuously
connected to EP1 as $\Omega$ passes through $\Omega_c^{(1)}$, the stability
structure across the EP is reversed: the region to the left of EP1 is
stable, whereas the region to the right of EP5 is stable. Thus, EP1 and
EP5 represent distinct types of stability boundaries despite being
continuously connected through the critical point.  Therefore, tuning the Raman coupling through
$\Omega_c^{(n)}$ not only reconstructs the exceptional-point structure but
also reverses the stability boundary associated with the EP. We refer to the structures
below and above $\Omega_c^{(n)}$ as the subcritical and supercritical EP
structures, respectively.
The same evolution occurs for all angular-momentum sectors $n$ with
$n<2l$, for which both subcritical and supercritical EP structures are
accessible. For $n=2l$, one has $\Omega_c^{(2l)}=0$, so any finite positive
Raman coupling places the system in the supercritical regime. For
$n>2l$, the subcritical regime is
inaccessible for $\Omega>0$; these sectors exhibit only the supercritical
EP structure.

\begin{figure}[tb]
    \centering
    \hspace*{-0.07\linewidth}
    \includegraphics[width=1.0\linewidth]{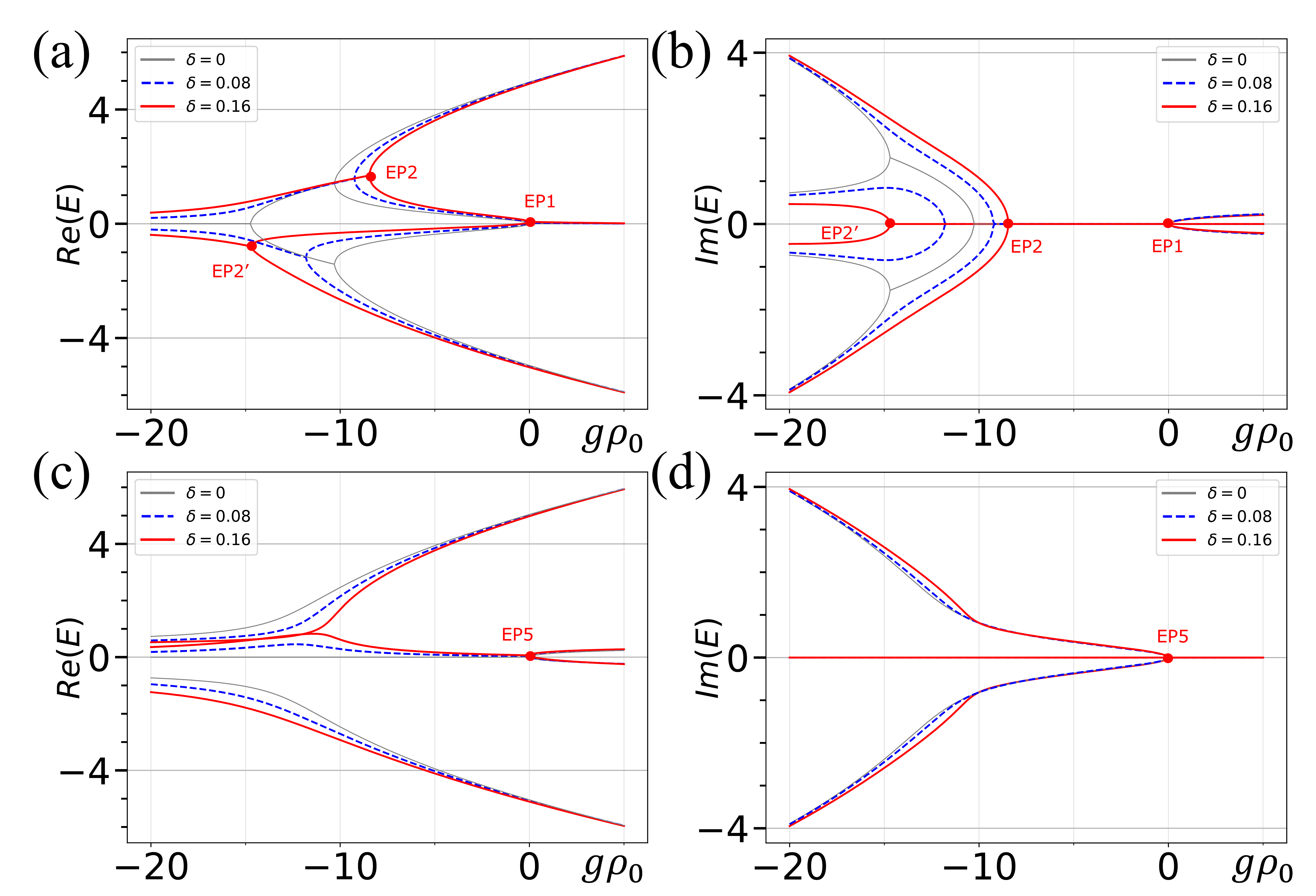}
    \caption{Effect of a finite two-photon detuning on the BdG spectrum for
$n=1$, $l=1$, and $m_{0}=0$.
(a) Real and (b) imaginary parts of the BdG eigenvalues for
$\Omega=2.95<\Omega_{c}$, with
$\delta=0$ (gray),
$\delta=0.08$ (blue dashed), and
$\delta=0.16$ (red solid).
(c) Real and (d) imaginary parts of the BdG eigenvalues for
$\Omega=3.05>\Omega_{c}$ with the same values of $\delta$.
For $\Omega<\Omega_{c}$, a finite detuning lifts the degeneracy
between opposite angular-momentum sectors and splits the
double-EP structure into two independent second-order EPs.}
    \label{fig:fig3}
\end{figure}

We next investigate the effect of the two-photon detuning $\delta$. At
$\delta=0$, the chiral symmetry leads to coincident symmetry-related EPs. A finite detuning lifts the single-particle energy degeneracy between $m$ and $-m$ and breaks this
chiral symmetry. Equivalently, the spectral constraint
$E\leftrightarrow-E$ is no longer imposed. Consequently, the EP2 double-EP
structure  splits into two distinct second-order EPs whose positions can be
controlled by the detuning, as illustrated in Fig.~\ref{fig:fig3}. It is worth noting that this
splitting does not affect the intrinsic particle-hole structure of the full
BdG Hamiltonian, which continues to enforce the spectral relation
$E\leftrightarrow-E^*$ for $H_{\rm BdG}$.
Figs.~\ref{fig:fig3}(a) and \ref{fig:fig3}(b) show the real and imaginary
parts of the dynamical spectrum for $l=1$, $m_0=0$ and $\Omega=2.95<\Omega_c^{(1)}$ with different
values of $\delta$. Increasing $\delta$ shifts EP2 and EP2$'$ in opposite
directions, while the EP1 position remains nearly unchanged for the weak
detuning considered here. The EP3 double-EP structure disappears once the chiral
symmetry is broken. For $\Omega=3.05>\Omega_c$, the EP5 structure persists
under finite detuning, although the detailed spectrum in the unstable regime
is modified, as shown in
Figs.~\ref{fig:fig3}(c) and \ref{fig:fig3}(d). The fourth-order EP and the associated exceptional line at
$\Omega=\Omega_c^{(n)}$ also disappear for finite $\delta$.

\begin{figure}[tb]
    \centering
    \hspace*{-0.07\linewidth}
    \includegraphics[width=0.95\linewidth]{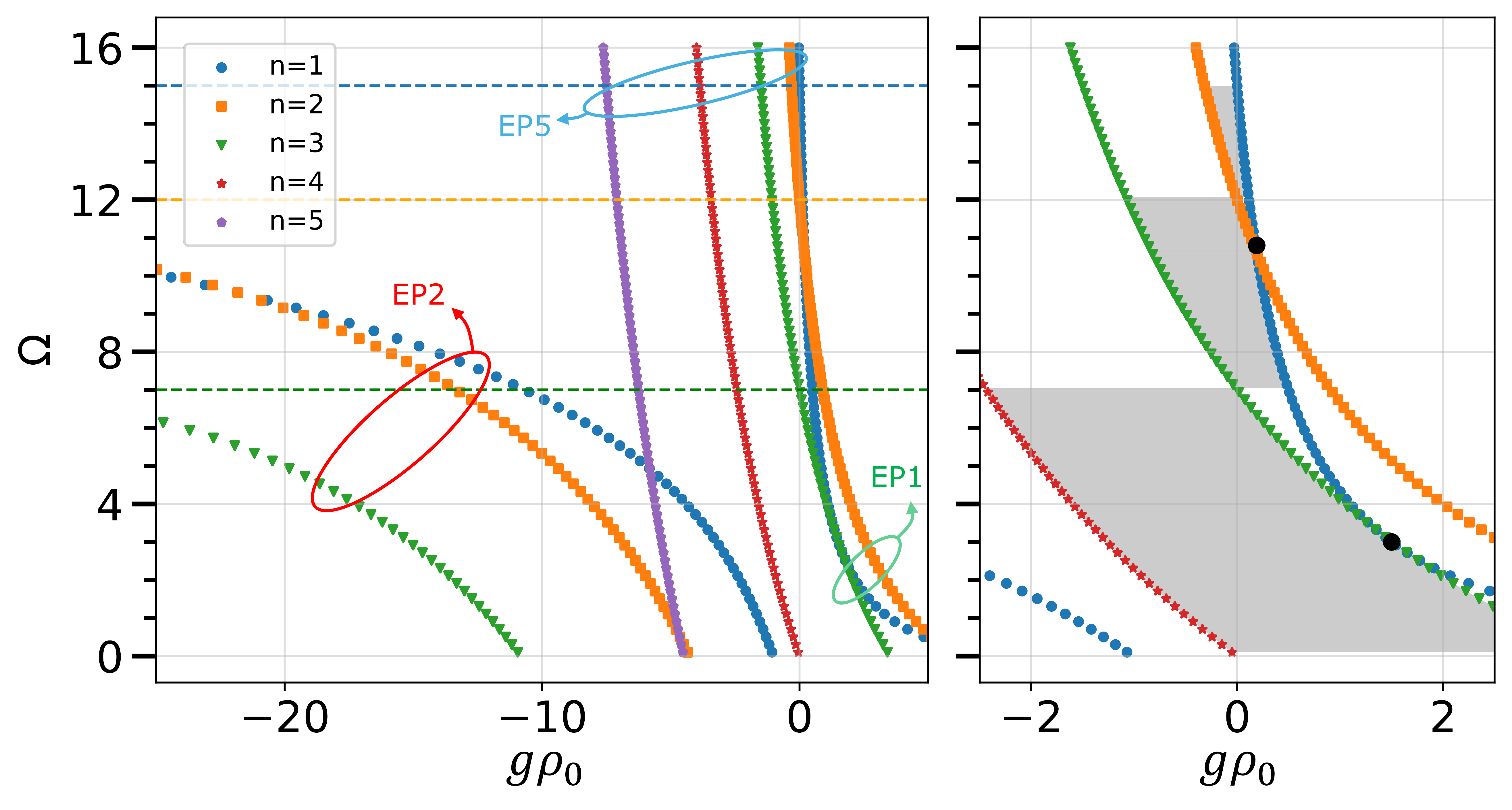}
    \caption{Exceptional-point landscape in the $(g,\Omega)$ parameter
space for $l=2$, $m_{0}=0$, and $\delta=0$.
The left panel shows the EP trajectories for the
angular-momentum sectors $n=1,\ldots,5$. The horizontal dashed lines indicate the corresponding critical Raman couplings.
The right panel enlarges the weak-interaction region relevant for
quantum sensing. The gray shaded region denotes the globally
stable parameter regime in which no 
angular-momentum sector has entered the PT-broken phase.
The black dots indicate intersections between EP trajectories
of different angular-momentum sectors and represent candidate
operating points for multimode sensing. }
    \label{fig:fig4}
\end{figure}

We now map the EP trajectories across different angular-momentum sectors to
obtain the global exceptional-point landscape. We first consider the
chiral-symmetric case $\delta=0$ and $m_0=0$ and plot the EP trajectories in
the $(g,\Omega)$ parameter plane for different sectors, as shown in the left
panel of Fig.~\ref{fig:fig4}. We consider $l=2$, the critical Raman couplings for
$n=1,2,3,4$ are $Omega_c^{(1)}=15$, 
$\Omega_c^{(2)}=12$, 
$\Omega_c^{(3)}=7$ 
and $\Omega_c^{(4)}=0$,
while the critical Raman coupling for $n=5$ does not exist. Accordingly, the $n=1,2,3$ sectors support both
subcritical and supercritical EP structures. As $\Omega$ increases across
the corresponding critical coupling, their EP1 trajectories approach
$g=0$ and continuously evolve into the EP5, and the stability properties change dramatically. In contrast, the $n=4,5$ sectors exhibit only the supercritical EP branch.

The multimode nature of the SOAM-coupled BEC introduces an additional global
stability constraint. A given excitation sector may remain dynamically stable
as it approaches its target EP, while another angular-momentum sector has
already developed complex BdG eigenvalues. We therefore distinguish the
sector-resolved stability of a target mode from the global stability of the
full BdG dynamics. For an EP to be accessible from a globally stable state,
all excitation sectors must remain dynamically stable before the target EP is
reached. The globally stable region satisfying this condition is shown by the
gray shaded area in the right panel of Fig.~\ref{fig:fig4}.

This global stability constraint strongly restricts the accessible EP
structures. In particular, the $n=2l$ sector has $\Omega_c^{(2l)}=0$ and therefore
remains on the supercritical side for any positive Raman coupling. As $\Omega$ increases from 0, the EP5 of sector $n=2l$ moves from $g=0$ toward negative $g$,  
so that this sector can become dynamically unstable before the EP structures of lower-angular-momentum sectors are reached as the interaction strength is varied.
Some EPs that are
well-defined within an individual sector are therefore inaccessible from a
globally stable state.
This demonstrates that the existence and order of an
EP alone do not determine its accessibility; the surrounding multimode
stability landscape must also be taken into account.

\begin{figure}[tb]
    \centering
    \hspace*{-0.07\linewidth}
    \includegraphics[width=0.95\linewidth]{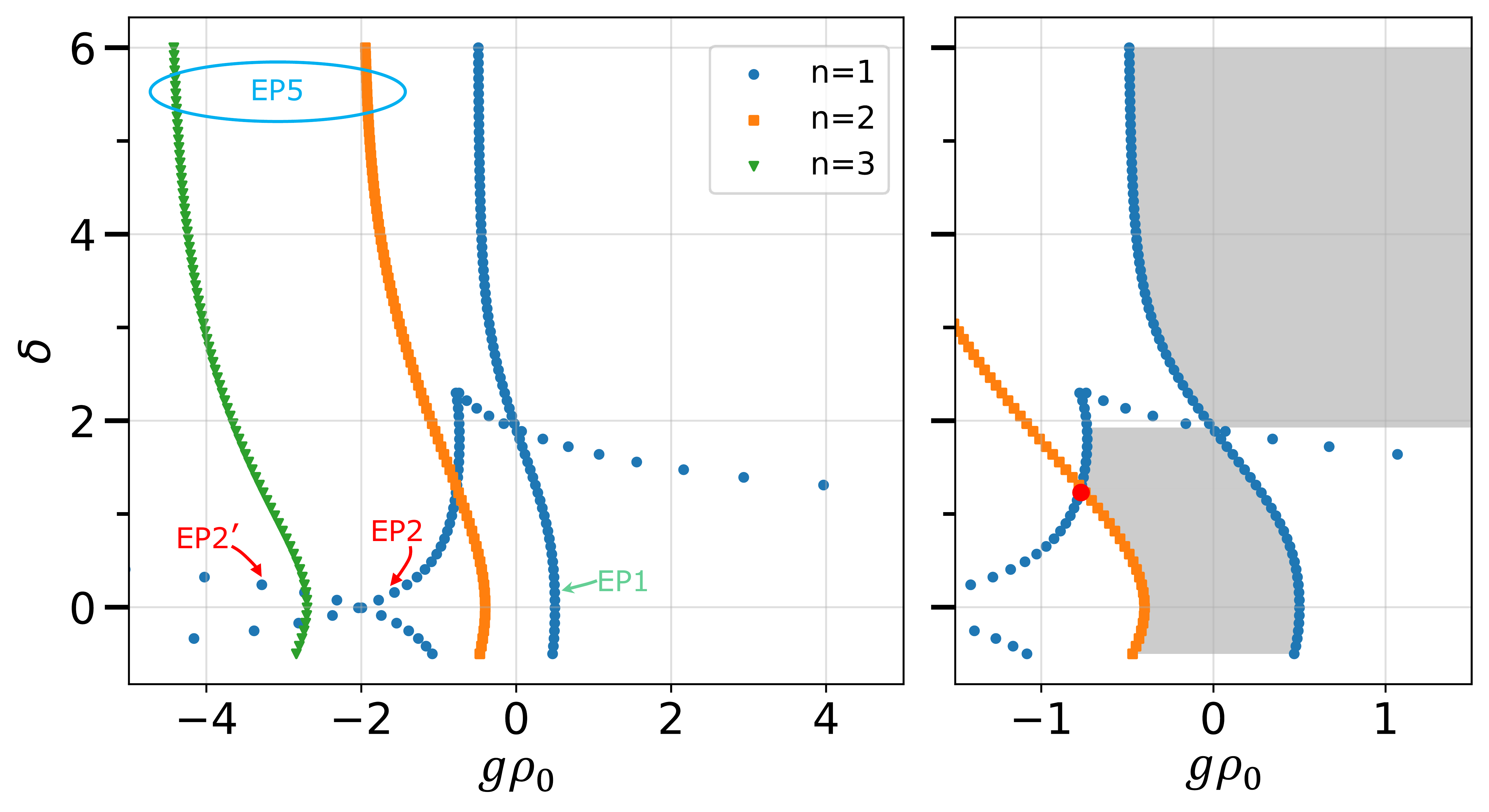}
    \caption{Exceptional-point landscape in the $(g,\delta)$ parameter space for $l=1$, $m_0=0$, and $\Omega=1$. The left panel shows the EP trajectories for the
angular-momentum sectors $n=1,2,3$. The right panel shows an enlarged view of the region containing globally stable parameters (gray shaded area) and the multimode EP intersection (red dot).}
    \label{fig:fig7}
\end{figure}

A similar tunability can be obtained by varying the two-photon detuning.
Fig.~\ref{fig:fig7} shows the EP trajectories in the $(g,\delta)$ parameter
plane for $l=1$, $m_0=0$, and $\Omega=1$. The different curves correspond to
EP trajectories associated with different angular-momentum sectors, while
the gray shaded region denotes the globally stable parameter regime. Compared
with the $(g,\Omega)$ phase diagram, varying $\delta$ provides an additional
degree of freedom for controlling the locations and intersections of EPs.
Interestingly, EP trajectories from different angular-momentum sectors can
also intersect. At such intersections, two independent excitation sectors
simultaneously reach their respective second-order EPs. These points,
indicated by the black dots in Fig.~\ref{fig:fig4} and the red dot in the right panel of Fig.~\ref{fig:fig7}, allow two independent EP channels to be activated
simultaneously and therefore provide natural
candidate operating points for multimode sensing which will be considered in the sensing analysis below.

\section{Quantum Sensing}

Having established the tunable exceptional-point landscape and the globally
accessible parameter regimes, we now investigate their metrological
consequences. Near an EP, the coalescence of eigenvalues and eigenvectors
gives rise to nontrivial dynamical amplification, which can enhance the
distinguishability of quantum states under small variations of a system
parameter. We quantify this enhancement using the QFI.

For a pure quantum state $\lvert\psi_\eta\rangle$ that depends on a parameter
$\eta$, the QFI is defined as
\begin{equation}
F_\eta
=
4\left[
\langle\partial_\eta\psi_\eta
|
\partial_\eta\psi_\eta\rangle
-
\left|
\langle\psi_\eta
|
\partial_\eta\psi_\eta\rangle
\right|^2
\right].
\label{eq:QFI-general}
\end{equation}
The QFI quantifies the distinguishability of quantum states generated by
nearby values of $\eta$ and sets the quantum Cram\'er--Rao bound on the
precision of unbiased parameter estimation. For the Bogoliubov dynamics
considered here, the QFI can be evaluated directly from the evolution
generated by the full BdG dynamical matrix $H_{\rm BdG}$. Following
Ref.~\cite{Liu2024QFIScalingQEP}, the QFI for a parameter $\eta$ entering
$H_{\rm BdG}$ can be written as
\begin{equation}
F_\eta
=
4B^\dagger B
+
2\operatorname{Tr}
\left(
C_2^\dagger C_2
\right),
\label{eq:QFI-BdG}
\end{equation}
with
$B=C_1\alpha+C_2\alpha^*$.
Here $[\alpha,\alpha^*]$ denotes the vector of initial expectation values of
the Bogoliubov operators $\hat V$. The matrices $C_1$ and $C_2$ characterize
the response of the BdG evolution to the parameter $\eta$ and are determined
from
\begin{equation}
\begin{pmatrix}
C_1 & C_2\\
C_2^* & C_1^*
\end{pmatrix}
=
\int_0^t dy\,
S^\dagger(y)\Sigma_z
\left(\partial_\eta H_{\rm BdG}\right)
S(y),
\label{eq:C1C2}
\end{equation}
where
\begin{equation}
S(y)=e^{-iyH_{\rm BdG}},
\qquad
\Sigma_z=\tau_z\otimes I_4,
\end{equation}
with $\tau_z$ acting in the particle-hole space. This formulation allows the
QFI to be evaluated directly from the BdG evolution without explicitly
constructing the time-evolved many-body wave function.

In the present system, the interaction strength $g$, Raman coupling
$\Omega$, and two-photon detuning $\delta$ can all serve as potential sensing
parameters. In the following, we focus primarily on the estimation of $g$,
with $\Omega$ considered as an additional control and sensing parameter where
appropriate. We do not use $\delta$ as the primary sensing parameter because
varying $\delta$ changes not only the BdG dynamical matrix but also the
condensate spinor and the corresponding mean-field background. By contrast,
for isotropic interactions, varying $g$ provides a direct control of the
interaction-induced Bogoliubov dynamics while leaving the condensate spinor
unchanged for a given operating point.

We first consider the sensing performance near an isolated second-order EP.
As established above, only EPs approached from the globally stable regime are
relevant for the present sensing protocol. We therefore initialize the system
in a stable state and vary the sensing parameter toward the target EP. The
resulting QFI exhibits a pronounced enhancement as the EP is approached,
reflecting the critical amplification of the Bogoliubov dynamics. The QFI
enhancement near an isolated EP is consistent with that found in
single-component BECs~\cite{Luo2022PseudoAPT}, while the SOAM coupling
provides substantially greater tunability of the EP location.

\begin{figure}[tb]
    \centering
    \hspace*{-0.07\linewidth}
    \includegraphics[width=1.1\linewidth]{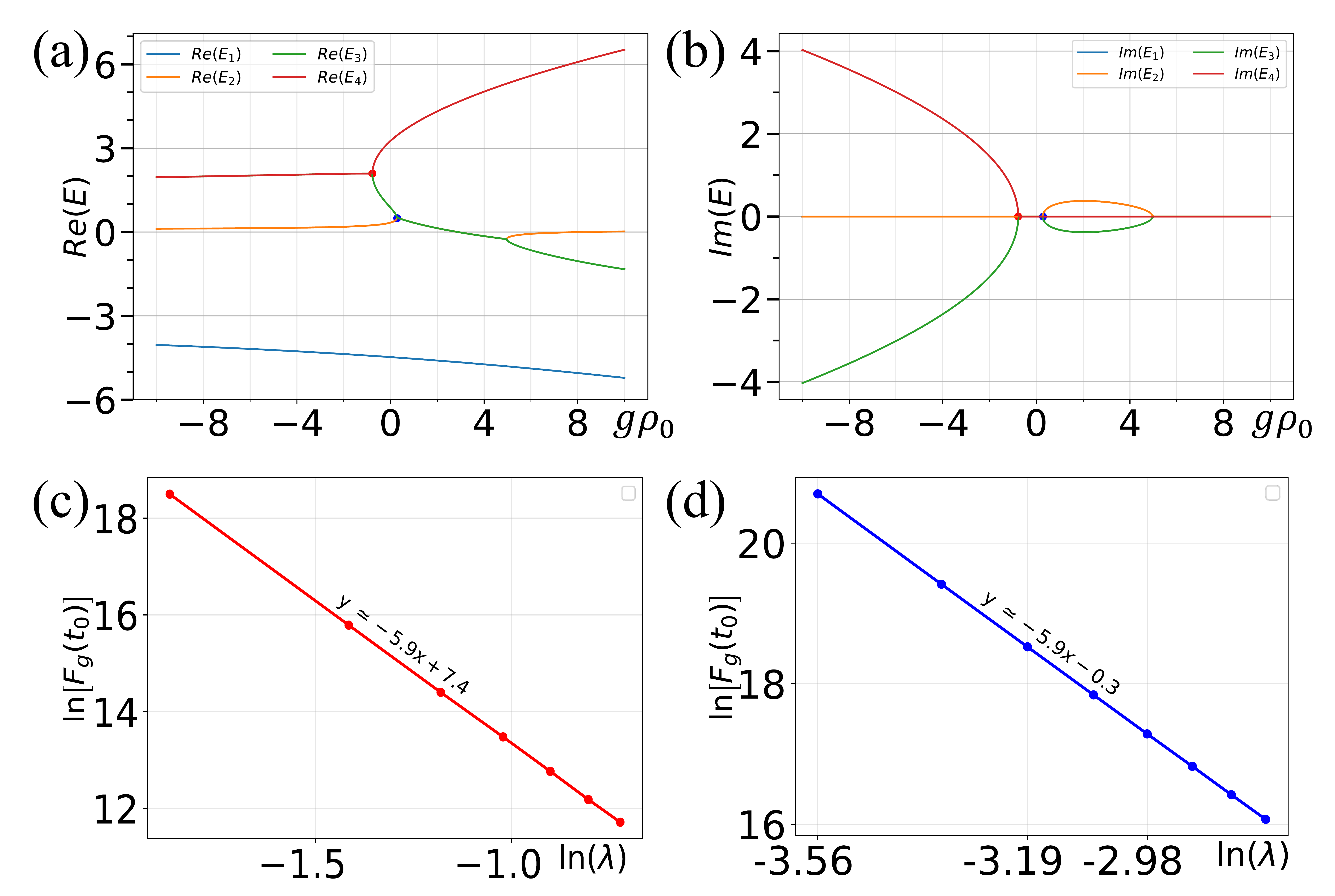}
    \caption{(a) Real and (b) imaginary parts of the BdG spectrum for
    $n=1$, $l=1$, $m_0=0$, $\delta=1.25$, and $\Omega=1$.
    (c) $\ln[F_g(t_0)]$ versus $\ln(\lambda)$ near the EP marked
    by the red dot in (a), as $g$ approaches the EP from the stable side. (d) $\ln[F_g(t_0)]$ versus $\ln(\lambda)$
    near the EP marked by the blue dot in (a), as $g$ approaches the EP from the stable side, with $\alpha=e^{-i\frac{\pi}{4}}(1,1,i,i)^T$.}
    \label{fig:fig6}
\end{figure}

To characterize the critical scaling, we introduce the quasiparticle energy
gap $\lambda$ associated with the target EP and evaluate the QFI at the
characteristic evolution time $t_0=\frac{\pi}{\lambda}$~\cite{Luo2022PseudoAPT}.
As the EP is approached from the stable side, $\lambda$ vanishes and the QFI
increases rapidly. For the second-order EPs considered here, we find
numerically that
\begin{equation}
F_g(t_0)\sim \lambda^{-6}.
\label{eq:FI-lambda-scaling}
\end{equation}
This scaling demonstrates the strong critical
enhancement of the QFI near the exceptional points and is in good agreement with that found in Ref.~\cite{Luo2022PseudoAPT}. As examples, we consider
$l=1$, $m_0=0$, $n=1$, $\delta=1.25$, and $\Omega=1$, where two second-order
EPs are encountered as the interaction strength $g$ is varied, as shown in
Figs.~\ref{fig:fig6}(a) and \ref{fig:fig6}(b). The corresponding QFI
evaluated at the characteristic time $t_0=\pi/\lambda$ as a function of the
quasiparticle energy gap $\lambda$ for the two EPs are shown in
Figs.~\ref{fig:fig6}(c) and \ref{fig:fig6}(d). Both exhibit similar critical
scaling, with fitted exponents of approximately $-5.9$ and $-5.8$,
respectively.

Although the critical scaling exponents are similar for different EPs, their
QFI curves can differ substantially in magnitude at the same energy-gap scale
$\lambda$ (corresponding to the same characteristic evolution time
$t_0=\pi/\lambda$). The difference is reflected in the prefactor of
\begin{equation}
F_g(t_0)\simeq a\lambda^{-6}.
\label{eq:FI-prefactor}
\end{equation}
The prefactor generally depends on both the eigenvalues and eigenstates of
$H_{\rm BdG}$ and their parameter dependence. If the dynamics near two EPs
are dominated by similar effective two-mode squeezing processes, the
corresponding eigenstate contributions may be similar, in which case the
eigenvalue response provides a natural factor distinguishing their sensing
performance. In particular, the Bogoliubov eigenvalue gap can open at
different rates as the interaction strength moves away from the two EPs.
A larger gap-opening rate indicates a stronger response of the eigenenergy
to the parameter variation and can therefore be associated with a larger
QFI enhancement. 
In the present example, we find that the EP with the larger gap-opening rate
indeed exhibits a larger QFI prefactor at the same energy-gap scale. Although
the gap-opening rate is not the sole factor determining the QFI, it can serve
as a useful relative indicator for comparing the sensing performance of EPs
within a fixed parameter configuration. This discussion shows that the EP
order alone does not determine the absolute sensing performance. Even EPs of
the same order within the same BdG spectrum can exhibit substantially
different metrological responses.

Beyond optimizing individual EP channels, the tunability of the present system also allows multiple sensing channels to be exploited simultaneously. We now consider the possibility of simultaneously exploiting exceptional
points from different angular-momentum sectors. As discussed above, the
tunable SOAM dispersion allows EP trajectories associated with different
Bogoliubov sectors to intersect in parameter space. At such an intersection,
two independent excitation sectors can simultaneously approach their
respective second-order EPs while the system remains within the globally
stable regime.
Within the Bogoliubov approximation, different angular-momentum sectors are
dynamically independent. At an intersection between the EP trajectories of
two sectors, the corresponding BdG dynamical matrix can therefore be written
as a direct sum,
\begin{equation}
H_{\rm BdG}^{(n_1\oplus n_2)}
=
H_{\rm BdG}^{(n_1)}
\oplus
H_{\rm BdG}^{(n_2)}.
\end{equation}
For an initially separable state with no correlations between the two
sectors, the QFI is additive,
\begin{equation}
F_{\eta}^{\rm tot}
=
F_{\eta}^{(n_1)}
+
F_{\eta}^{(n_2)}.
\end{equation}
More generally, one has $F_{\eta}^{\rm tot}=\sum_n F_{\eta}^{(n)}$.
Thus, multiple independent EP channels can contribute simultaneously to the
total QFI without requiring a higher-order EP.

\begin{figure}[tb]
    \centering
    \hspace*{-0.07\linewidth}
    \includegraphics[width=1.1\linewidth]{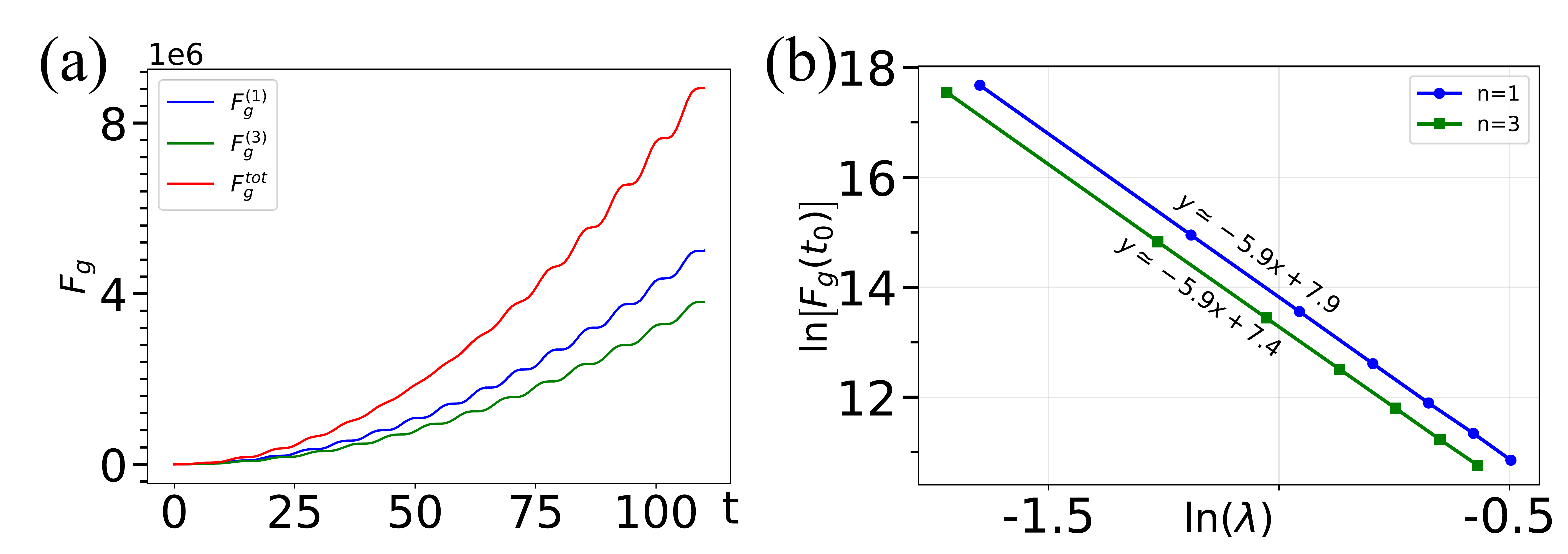}
    \caption{Multimode enhancement of the QFI for the
    lower intersection point indicated by the black dot in Fig.~\ref{fig:fig4} with $\Omega=3$ and
    $\alpha=e^{-i\frac{\pi}{4}}(1,1,i,i)^T$. (a) Time evolution of the quantum Fisher
    information contributed by the $n=1$ (blue) and $n=3$ (green)
    Bogoliubov sectors. The red curve shows the total QFI,
    $F_g^{\mathrm{tot}}=F_g^{(1)}+F_g^{(3)}$,
    demonstrating the additive enhancement produced by two dynamically
    independent EP channels, with $g\rho_0=1.4$. (b) Log--log plots of
    $F_g(t_0)$ as a function of the quasiparticle energy gap $\lambda$ for
    the two sectors, where $t_0=\pi/\lambda$. Linear fits yield slopes close
    to $-5.9$, consistent with the critical scaling
    $F_g(t_0)\sim\lambda^{-6}$. Other parameters are the same as that in Fig.~\ref{fig:fig4}.}
    \label{fig:fig5}
\end{figure}

As a representative example, we consider the lower EP intersection shown in
Fig.~\ref{fig:fig4}, where the EP trajectories of the $n=1$ and $n=3$
sectors intersect. At this operating point, both sectors simultaneously
approach their respective second-order EPs while the full system remains
globally stable. Fig.~\ref{fig:fig5}(a) shows the QFI for estimating $g$
from the two sectors and their total contribution. The two sector
contributions add independently, resulting in a substantially larger total
QFI than that obtained from either EP channel alone.

Because the two sectors have different characteristic energy gaps, their QFI
responses evolve differently with time, while the total QFI follows directly
from their sum. Fig.~\ref{fig:fig5}(b) further shows that both EP channels
retain the same critical scaling $F_g\left(t_0\right)
\sim \lambda^{-6}$,
with fitted exponent $\sim-5.9$. Thus, the multimode enhancement does not
arise from a modified critical exponent, but from the additive contributions
of independent EP channels.
This multimode mechanism provides an additional sensing resource enabled by
the SOAM-coupled condensate: multiple second-order EP channels can be
activated simultaneously, yielding a linear enhancement of the total QFI
without requiring the engineering of a single higher-order EP.

\section{Measurement Scheme}

In the previous section, we characterized the parameter sensitivity near
exceptional points in terms of the QFI, which
sets the ultimate precision allowed by quantum mechanics. Reaching this
bound, however, generally requires an optimal measurement that may not be
directly accessible experimentally. We therefore ask whether the
EP-enhanced sensitivity can be extracted using experimentally feasible
observables. In this section, we construct a density-based measurement
protocol and show that a substantial fraction of the QFI can be accessed
through experimentally accessible density measurements.

The quantum Cram\'er-Rao bound gives
\begin{equation}
(\Delta\eta)^2\geq \frac{1}{F_\eta},
\end{equation}
where $F_\eta$ is the QFI of the parameter-dependent quantum state. For a
specific observable $\hat O$, the corresponding sensitivity obtained from
error propagation is
\begin{equation}
\mathcal{S}_{\eta}
=
\frac{
\left|\partial_\eta\langle\hat O\rangle\right|^2
}{
(\Delta O)^2
}=(\Delta \eta)^{-2}
\label{eq:measurement_sensitivity}
\end{equation}
with $(\Delta O)^2=\langle\hat O^2\rangle-\langle\hat O\rangle^2$, which necessarily satisfies $\mathcal{S}_{\eta}\leq F_\eta$.
Thus, the ratio between the sensitivity extracted from a realistic
observable and the QFI directly characterizes how efficiently the
measurement accesses the metrological information generated by the
EP-enhanced dynamics.

Within the Bogoliubov framework, the enhanced sensitivity originates from
the amplified quasiparticle fluctuations associated with the EP dynamics. A
natural strategy is therefore to probe observables that are directly
sensitive to these fluctuations. In the ring-BEC system, Bogoliubov
excitations generate spatial modulations of the atomic density.
Spin-resolved density distributions are experimentally accessible through
standard imaging techniques and provide a direct probe of the excited
angular-momentum modes. We therefore start from the spin-resolved density
operators
\begin{equation}
\hat{\rho}_{s}(\phi)
=
\hat{\Psi}_{s}^{\dagger}(\phi)
\hat{\Psi}_{s}(\phi),
\end{equation}
and construct a density-based measurement operator from their appropriate
combination. For the present system, we find that the total-density channel
is sufficient to capture the enhanced sensing signal. We therefore define
\begin{equation}
\hat{\rho}(\phi)
=
\hat{\rho}_{\uparrow}(\phi)
+
\hat{\rho}_{\downarrow}(\phi),
\end{equation}
and construct the measurement operator
\begin{equation}
\hat O^{(n)}(\theta)
=
\int_{0}^{2\pi}
d\phi\,
\cos(n\phi+\theta)
\hat{\rho}(\phi),
\label{eq:density_measurement}
\end{equation}
where $\theta$ selects the spatial quadrature of the density modulation. The role of $\theta$ can be made explicit by writing
$\cos(n\phi+\theta)=\cos (n\phi)\cos\theta-\sin(n\phi)\sin\theta$ and
defining the two density quadratures
\begin{align}
\hat X^{(n)}
&=
\int_{0}^{2\pi}
d\phi\,
\cos (n\phi)\,\hat{\rho}(\phi),
\\
\hat Y^{(n)}
&=
\int_{0}^{2\pi}
d\phi\,
\sin(n\phi)\,\hat{\rho}(\phi),
\end{align}
the measurement operator becomes
\begin{equation}
\hat O^{(n)}(\theta)
=
\hat X^{(n)}\cos\theta-\hat Y^{(n)}\sin\theta.
\end{equation}
Thus, varying $\theta$ selects the spatial quadrature carrying the
parameter-dependent density signal without modifying the underlying
dynamics. We optimize this quadrature by maximizing $\mathcal{S}_{\eta}^{(n)}(\theta)$
at the chosen sensing point. In practice, we find that $\theta=0$ lies
very close to the optimal value and provides nearly optimal sensitivity.
We therefore set $\theta=0$ in the numerical simulations below.
Therefore, the measurement operator reduces to $X^{(n)}$ which can be derived as
\begin{equation}
\hat X^{(n)}=\sqrt{\frac{\pi}{2}}(\hat A+\hat A^\dagger)
\end{equation}
with
\[\hat A=
e^{-i\frac{\pi}{4}}
\begin{pmatrix}
\Phi_\uparrow &
\Phi_\downarrow &
i\Phi_\uparrow &
i\Phi_\downarrow
\end{pmatrix}
\begin{pmatrix}
\hat\psi_{m\uparrow}\\
\hat\psi_{m\downarrow}\\
\hat\psi^\dagger_{m'\uparrow}\\
\hat\psi^\dagger_{m'\downarrow}
\end{pmatrix}\]
and $m,m'=m_0\pm n$.

\begin{figure}[tb]
\centering
\hspace*{-0.07\linewidth}
\includegraphics[width=1.1\linewidth]{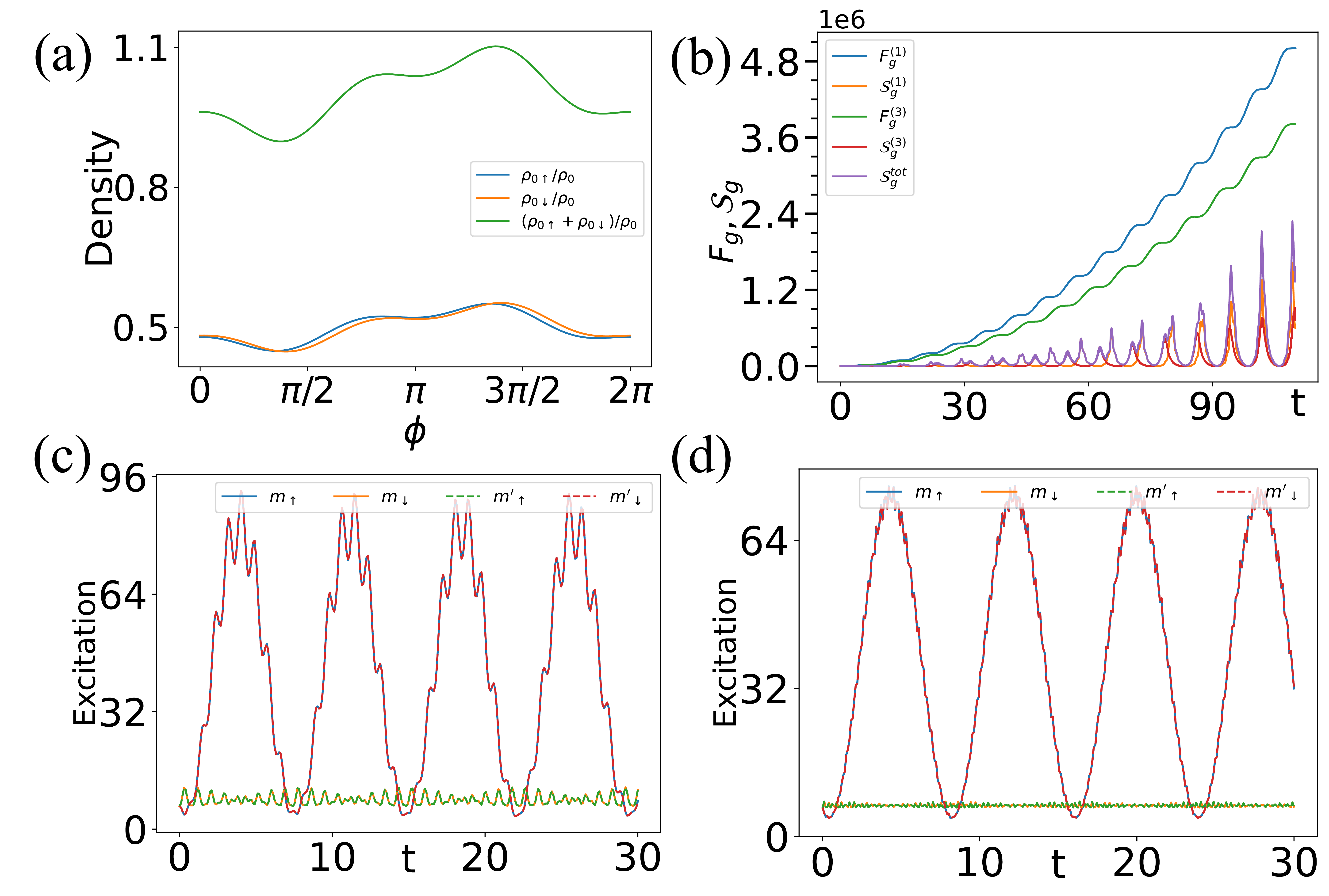}
\caption{(a) Normalized spin-resolved densities together with the normalized total-density, with $t=120$. (b) Time evolution of the QFI and the measurement sensitivities based on the density observable for the $n=1$ and $n=3$ sectors, together with the total measurement sensitivities. (c) and (d) Time evolution of the excitation populations of the two excitation sectors $n=1,3$, respectively, with $m=m_0+n$, $m'=m_0-n$. In all plot, $g\rho_0=1.4$ and other parameters are the same as that in Fig.~\ref{fig:fig5}.}
\label{fig:fig8}
\end{figure}

We consider a ring-BEC system with condensate atom number $N_0=6000$, the condensate density is $\rho_0=N_0/2\pi$.
We tune the parameter to the the multimode sensing configuration discussed above, where
the $n=1$ and $n=3$ Bogoliubov sectors provide independent sensing channels. The typical spin-revolved density $\langle\hat\rho_s\rangle/\rho_0$ and total density $\langle\hat\rho\rangle/\rho_0$ are shown in Fig.~\ref{fig:fig8}(a). From which we can extract the value of measurement operator
$\langle \hat X^{(n)}\rangle$.
In Fig.~\ref{fig:fig8}(b), we consider the sensing parameter $\eta=g$ and plot the simulated QFI and the measurement sensitivity as functions of evolution time. For both excitation sectors, the measurement sensitivities are of the
same order as the corresponding QFI, demonstrating that a substantial
fraction of the metrological information can be extracted from the density
observable.
For independently resolved channels, the measurement information can likewise be combined. If $\mathcal{S}_{\eta}^{(n)}$ denotes the
error-propagation sensitivity obtained from the measurement of the $n$th
channel, the sensitivities of independent estimators can be combined at the
level of inverse variances
\begin{equation}
\mathcal{S}_{\eta}^{\mathrm{tot}}
=
\mathcal{S}_{\eta}^{(1)}
+
\mathcal{S}_{\eta}^{(3)}.
\label{eq:MA_addition}
\end{equation}
Thus, the metrological information carried by multiple independent
Bogoliubov channels can be accumulated not only at the QFI level, but also
through experimentally accessible density measurements.
It is worth noting that the energy gaps $\lambda$ are different for the two
sectors at the sensing point. Consequently, their characteristic enhancement
times, approximately given by
$t=\pi/\lambda,2\pi/\lambda,3\pi/\lambda,\ldots$,
do not necessarily coincide. Nevertheless, the two measurement channels exhibit overlapping sensitivity-enhancement windows during the evolution. When the enhancement windows of the two channels occur at different times, the total sensitivity is dominated by the channel with the stronger instantaneous response. In contrast, when the enhancement windows overlap, both channels contribute simultaneously and their independent measurement information adds, resulting in a pronounced enhancement of the total sensitivity. This multimode enhancement provides an experimentally accessible counterpart of the additive QFI enhancement identified above.

Finally, we verify the Bogoliubov description by examining the populations of the relevant
Bogoliubov excitation modes during the sensing dynamics, the results for the two sectors are shown in Figs.~\ref{fig:fig8}(c) and \ref{fig:fig8}(d), respectively. The excitation
numbers remain on the order of only a few tens of particles throughout the evolution considered here. Compared with the macroscopic condensate population, this corresponds to a negligible depletion fraction. The system
therefore remains within the weak-depletion regime required for the Bogoliubov approximation, ensuring that the mean-field background remains self-consistent throughout the sensing dynamics. The enhanced measurement sensitivity is thus obtained while remaining well within the weak-depletion regime.

\section{Conclusion}

In this work, we have developed a theoretical framework for tunable exceptional-point quantum sensing in a SOAM-coupled
BEC confined in a ring. Starting from a fully Hermitian
microscopic Hamiltonian, we derived the Bogoliubov--de Gennes dynamical matrix and showed that the SOAM-induced multimode structure supports multiple exceptional points, including second- and fourth-order EPs, whose locations can be tuned through the Raman coupling and interaction parameters. The resulting tunability provides access to a rich EP landscape with distinct stability regimes and excitation channels.

We further investigated the metrological consequences of these tunable EPs using the QFI. Approaching a second-order EP from the globally stable regime leads to strong critical enhancement of the QFI, with
the characteristic scaling $F_g(\pi/\lambda)\propto\lambda^{-6}$ found for the EPs considered here. We also find that EP order alone does not determine the sensing performance: EPs of the same order can exhibit substantially different QFI enhancements, with the gap-opening rate providing a useful relative indicator for comparing EPs within a fixed parameter configuration. Moreover, the multimode structure of the SOAM-coupled condensate allows independent second-order EP channels to be activated
simultaneously. Their QFI contributions add, providing a linear enhancement of the total metrological information without requiring a single higher-order EP.

Finally, we demonstrated that the EP-enhanced metrological information can be accessed through an experimentally feasible density-based measurement. By selecting an appropriate spatial density quadrature, the measurement sensitivity can approach the QFI for the individual EP channels. When the sensitivity-enhancement windows of different channels overlap, their measurement sensitivities can be combined to yield a substantial multimode enhancement. The excitation populations remain small compared with the condensate population throughout the sensing dynamics, ensuring the self-consistency of the Bogoliubov description. Overall, the SOAM-coupled ring BEC provides an intrinsic non-Hermitian platform for tunable EP-enhanced quantum sensing without engineered gain or loss. The controllable Raman coupling, interaction strength, and angular-momentum degrees of freedom offer flexible control over EP locations, stability, and multimode sensing channels, opening a route toward experimentally accessible quantum sensing based on tunable exceptional-point dynamics.

{\textit{Acknowledgments}.---}This work is supported by the National Natural Science Foundation of China (Grants No. 12574544), Quantum Science and Technology-National Science and Technology Major Project (Grant No. 2021ZD0301200), the CAS Project for Young Scientists in Basic Research (YSBR-085), and
USTC start-up funding.

{\textit{Data availability.---}}The data that support the findings of this study are publicly available~\cite{data}.


%


\end{document}